\documentclass[smallextended]{svjour3}
\usepackage{latexsym,amsmath,amscd,amssymb,graphicx,amsbsy,xcolor,color,amsfonts,mathptmx}

\begin{document}

\title{A geometric relativistic dynamics under any conservative force}

\author{Yaakov Friedman \and Tzvi Scarr \and Joseph Steiner}
\institute{Yaakov Friedman \at Jerusalem College of Technology, Departments of Mathematics and Physics, P.O.B. 16031 Jerusalem 91160, ISRAEL
   \\ Tel.: +972-2-675-1184, Fax: +972-2-675-1045  \\ \email{friedman@jct.ac.il} \and Tzvi Scarr \at Jerusalem College of Technology, Department of Mathematics, P.O.B. 16031 Jerusalem 91160, ISRAEL
     \\ Tel.: +972-2-675-1274, Fax: +972-2-675-1285  \\ \email{scarr@g.jct.ac.il}
     \and Joseph Steiner \at Jerusalem College of Technology, Departments of Mathematics and Physics, P.O.B. 16031 Jerusalem 91160, ISRAEL
   \\ Tel.: +972-2-675-1265  \\ \email {steiner@jct.ac.il} }

\date{Received: date / Accepted: date}

\maketitle

\begin{abstract}
Riemann's principle ``force equals geometry" provided the basis for Einstein's General Relativity - the geometric  theory of gravitation. In this paper, we follow this principle to derive the dynamics for any static, conservative force.  The geometry of spacetime of a moving object is described by a metric obtained from the potential of the force field acting on it.  We introduce a generalization of Newton's First Law - the \emph{Generalized Principle of Inertia} stating that: \emph{An inanimate object moves inertially, that is, with constant velocity, in \emph{its own} spacetime whose geometry is determined by the forces affecting it}.  Classical Newtonian dynamics is treated within this framework, using a properly defined \emph{Newtonian metric} with respect to an inertial lab frame. We reveal a physical deficiency of this metric (responsible for the inability of Newtonian dynamics to account for relativistic behavior), and remove it. The dynamics defined by the corrected Newtonian metric leads to a new \emph{Relativistic Newtonian Dynamics} for both massive objects and massless particles moving in any static, conservative force field, not necessarily gravitational. This dynamics reduces in the weak field, low velocity limit to classical Newtonian dynamics and also exactly reproduces the classical tests of General Relativity, as well as the post-Keplerian precession of binaries.
\vskip0.2cm
\end{abstract}

\spnewtheorem{theo}{Theorem}{\bf}{\rm}

\section{Introduction}\label{intro}
$\;$

Bernhard Riemann, although best known as a mathematician, became interested in physics in his early twenties. His lifelong dream was to develop the mathematics to unify the laws of electricity, magnetism, light and gravitation.
At an 1894 conference in Vienna, the mathematician Felix Klein said:
\begin{quotation}
``I must mention, first of all, that Riemann devoted much time and thought
to physical considerations. Grown up under the tradition which is represented
by the combinations of the names of Gauss and Wilhelm Weber,
influenced on the other hand by Herbart's philosophy, he endeavored
again and again to find a general mathematical formulation for the laws
underlying all natural phenomena .... The point to which I wish to
call your attention is that these physical views are the mainspring of
Riemann's purely mathematical investigations \cite{Klein}."
\end{quotation}

Riemann's approach to physics was geometric. As pointed out in \cite{papa}, ``one of the main features of the
local geometry conceived by Riemann is that it is well suited to the study of gravity
and more general fields in physics." He believed that the forces at play in a system determine the geometry of the system. For Riemann, \emph{force equals geometry}.

The application of Riemann's mathematics to physics would have to wait for two more essential ideas. While Riemann considered how forces affect \emph{space}, physics must be carried out in \emph{spacetime}. One must consider trajectories in spacetime, not in space. For example, in flat spacetime, an object moves with constant velocity if and only if his trajectory in \emph{spacetime} is a straight line. On the other hand, knowing that an object moves along a straight line in \emph{space} tells one nothing about whether the object is accelerating. As Minkowski said, ``Henceforth, space by itself, and time by itself, are doomed to fade away into mere shadows, and only a kind of union of the two will preserve an independent reality \cite{Mink}." This led to the second idea. Riemann worked only with positive definite metrics, whereas Minkowski's metric on spacetime is not positive definite. The relaxing of the requirement of positive-definiteness to non-degeneracy led to the development of \emph{pseudo-Riemannian geometry}.

Fifty years after Riemann's death, Einstein used pseudo-Riemannian geometry as the cornerstone of General Relativity ($GR$). Acknowledging his reliance on Riemann, Einstein said:
\begin{quotation}
``But the physicists were still far removed from such a way of thinking;
space was still, for them, a rigid, homogeneous something, incapable of
changing or assuming various states. Only the genius of Riemann, solitary
and uncomprehended, had already won its way to a new conception
of space, in which space was deprived of its rigidity, and the possibility
of its partaking in physical events was recognized. This intellectual
achievement commands our admiration all the more for having preceded
Faraday's and Maxwell's field theory of electricity \cite{Ein}."
\end{quotation}

$GR$ is a direct application of ``force equals geometry." In $GR$, the gravitational force curves spacetime. Since, by the Equivalence Principle, the acceleration of an object in a gravitational field is independent of its mass, curved spacetime can be considered a \emph{stage} on which objects move. In other words, the geometry is the same for all objects. However, the Equivalence Principle holds only for gravitation. In this way, $GR$ singles out the gravitational force from other forces which are not treated geometrically. For example, the potential of an electric force depends on the charge of the particle, and the particle's acceleration depends on its charge-to-mass ratio. Thus, the electric field does not create a common stage on which all particles move. Indeed, a neutral particle does not feel any electric force at all. The way spacetime curves due to an electric potential depends on \emph{both} the potential \emph{and} intrinsic properties of the object. This was also recognized in the geometric approach of \cite{Duarte}. How, then, are we to apply Riemann's principle of ``force equals geometry" to other forces?

In this paper, we realize Riemann's program for motion in any static, conservative force field. We describe the geometry of the spacetime of a moving object via a metric derived from the potential of the force field acting on the object. This metric is computed  with respect to an inertial lab frame.  The dynamics based on this geometry follows from our new \emph{Generalized Principle of Inertia} - a generalization of Newton's First Law. Classical Newtonian dynamics can be treated within this framework by properly defining a \emph{Newtonian metric}. We discover a physical deficiency of this metric which explains why Newtonian dynamics cannot account for relativistic behavior. We remove this deficiency and obtain a corrected Newtonian metric. The ensuing dynamics, called \emph{Relativistic Newtonian Dynamics} ($RND$), is applicable to both massive objects and massless particles.  This dynamics reduces in the weak field, low velocity limit to classical Newtonian dynamics and also exactly reproduces the tests of $GR$. We plan to extend the theory presented here to non-static forces using Lorentz covariance .

In the literature, there are other alternative approaches to reproducing the relativistic gravitational features of $GR$. One approach uses modified Newtonian-like potentials. This so-called ``pseudo-Newtonian" approach, introduced in \cite{PW}, is much simpler mathematically than $GR$, with no need for covariant differentiation and complicated tensorial equations. Numerous authors (\cite{Ab,Chak,Haw,Lee,Mac,Mat,Ross}) have proposed various modified Newtonian-like potentials.  However, none of these potentials are able to reproduce the tests of $GR$, even in the weak field regime. Moreover, as stated in \cite{Ghosh14}, most of these modified potentials ``are arbitrarily proposed in an ad hoc way" and, more fundamentally, are ``not a physical analogue of local gravity and are not based on any robust physical theory and do not satisfy Poisson's equation."

More recently, the above shortcomings were addressed in \cite{Ghosh16}. Using a metric approach and hypothesizing a generic relativistic gravitational action and a corresponding Lagrangian, the authors derive a velocity-dependent relativistic potential which generalizes the classical Newtonian potential. For a static, spherically symmetric geometry, this potential exactly reproduces relativistic gravitational features, including the tests of $GR$.
Even more recently, one finds a fundamental grounding to these velocity-dependent pseudo-Newtonian potentials in \cite{Lamm}. The authors generalize the pseudo-Newtonian approach to any stationary spacetime. They also include additional forces, such as the electromagnetic force.

The paper is organized as follows. In section \ref{gd}, we introduce the Generalized Principle of Inertia. Then we prove that under certain conditions, conjugate momenta are conserved along the trajectory of an object. In section \ref{gnd}, we derive a dimensionless energy conservation equation and use it to construct a \emph{Newtonian metric}. In Section \ref{refnewmet}, we analyze our Newtonian metric and discover a physical deficiency. We remove this deficiency and obtain a corrected Newtonian metric. For a spherically symmetric gravitational potential, the corrected metric is the Schwarzschild metric. The dynamics based on the corrected metric is $RND$. This dynamics reduces in the weak field, low velocity limit to classical Newtonian dynamics. In section \ref{rnd}, we derive the $RND$ energy conservation equation and the $RND$ equation of motion for both massive objects and massless particles. In section \ref{tests}, we show that $RND$ exactly reproduces the classical tests of $GR$.

$RND$ has the following features:
\begin{enumerate}
  \item It is based on the classical, unmodified Newtonian potential.
  \item It avoids the complicated field equations of $GR$.
  \item It reveals the physical mechanism responsible for relativistic phenomena.
  \item It is also valid for non-spherically symmetric fields.
  \item It does not rely on the Equivalence Principle, and so is applicable to \emph{any combination} of static, conservative force fields whose potentials vanish at infinity.
  \end{enumerate}

\section{A geometric approach to dynamics}\label{gd}

One of the main new ideas we present here is the \emph{relativity of spacetime}. By this, we mean that spacetime is an object-dependent notion. An object lives in its own spacetime, its own geometric world, which is defined by the forces
which affect it. For example, in the vicinity of an electric field, a charged particle and a neutral particle exist in different worlds, in different spacetimes. In fact, for the neutral particle, the electric field does not exist. Likewise, in the vicinity of a magnet, a piece of iron and a piece of plastic live in two different worlds.

An inanimate object has no internal mechanism with which to change its velocity. Hence, it has constant velocity in \emph{its own} world (spacetime). This leads us to formulate a new principle, the \emph{Generalized Principle of Inertia}, which generalizes Newton's First Law and states that:  \textbf{An inanimate object moves inertially, that is, with constant velocity, in \emph{its own} spacetime whose geometry is determined by the forces affecting it}. This is a generalization, or more accurately, a relativization of what Einstein accomplished. In $GR$, an object freely falling in a gravitational field is in free motion. This is attested to by the fact that along a geodesic, the acceleration is zero. The Generalized Principle of Inertia states that \emph{every} object is in free motion in \emph{its} spacetime, determined by the  forces affecting  it.

An object obeying Newton's First Law moves with constant velocity. Its trajectory is a straight line in spacetime. Moreover, the distance between any two points on the trajectory is extremal  among all paths connecting these two points. Since, by the Generalized Principle of Inertia, an object moves with constant velocity in its own spacetime, we assume that there exists a metric with respect to which the length of the object's trajectory is extremal. This metric will depend only on the forces, and, in the case of static, conservative forces, the metric will depend only on the potentials of these forces. We call this metric the \emph{metric of the object's spacetime}.

Since we require a metric which will extremize the length of trajectories, we will use a variational principle and the ensuing conservation properties.
\begin{quotation}
``Many results in both classical and quantum physics can be expressed as variational principles, and it is often when expressed in this form that their physical
meaning is most clearly understood. Moreover, once a physical phenomenon has been written as a variational principle, ... it is usually possible to identify conserved
quantities, or symmetries of the system of interest, that otherwise might be found only with considerable effort \cite{Riley}."
\end{quotation}

Let $q:\sigma \to x, a\le\sigma\le b$ be a trajectory of an object, where $\sigma$ is an arbitrary parameter. Let
\begin{equation}\label{arbmet}
ds^2=g_{ij}(q)dq^idq^j
\end{equation}
be the metric of the object's spacetime, where $q^\alpha, \alpha=0,..,3$ are the coordinates in an inertial frame $K$ far removed from the sources of the field. The choice of $K$ corresponds to the ``rest frame of the universe" as in \cite{Ni}. Define
\begin{equation}\label{L}
L\left(q,\acute{q}\right)=\frac{ds}{d\sigma}=\sqrt{g_{ij}(q)\acute{q}^i\acute{q}^j},
\end{equation}
where $\acute{q}=\frac{dq}{d\sigma}$. The length $l(q)$ of the trajectory $q$ is given by
\begin{equation}\label{lq}
l(q)=\int_a^b \frac{ds}{d\sigma}d\sigma=\int_a^b L\left(q,\acute{q}\right)d\sigma.
\end{equation}
It is well known that the length of the trajectory does not depend on the parametrization.

Let $u:\sigma \to x, a\le\sigma\le b, u(a)=u(b)=0$ be a perturbation of the trajectory. The length of $q$ is extremal if
\begin{equation}\label{extcon}
\frac{d}{d\epsilon}l(q+\epsilon u)|_{\epsilon=0}=0.
\end{equation}

From the Generalized Principle of Inertia, the trajectory $q$ is extremal. By a standard argument it follows that the trajectory satisfies the Euler-Lagrange equations
\begin{equation}\label{EL eqns}
\frac{\partial L}{\partial q^i}-\frac{d}{d\sigma}\frac{\partial L}{\partial \acute{q}^i}=0,
\end{equation}
with $L$ defined by (\ref{L}).

For each coordinate $q^i$, we define the $i$-th component $p_i$ of the conjugate momentum by
\begin{equation}\label{defmomentum}
p_i = \frac{\partial L}{\partial\acute{q}^i}.
\end{equation}
For $L$ defined by  (\ref{L}), we have
\begin{equation}\label{compMoment}
  p_i = \frac{\partial L}{\partial\acute{q}^i}=\frac{g_{ij}\acute{q}^j}{\sqrt{g_{ij}(q)\acute{q}^i\acute{q}^j}}=\frac{g_{ij}\acute{q}^j}{ds/d\sigma}=g_{ij}\frac{dq^j}{ds}.
\end{equation}

Note that the second term in equation (\ref{EL eqns}) contains differentiation by two parameters on the curve. The first differentiation is by $s$, as seen in equation (\ref{compMoment}). The second differentiation is by $\sigma$. In order to obtain a differential equation with a single parameter, we will choose $\sigma$ to be proportional to $s$. More precisely, we choose $\sigma$ to be the parameter $\tau=c^{-1}s$, called \emph{proper time}, which is proportional to $s$ and reduces to the coordinate time $t$ in the classical limit. Using $\tau$ will turn equation (\ref{EL eqns}) into a second-order differential equation. We denote differentiation of $q$ with respect to $\tau$ by $\dot{q}$.

The following proposition follows immediately from equation (\ref{EL eqns}).

\noindent\textbf{Proposition 1}\;\;\; If the metric coefficients $g_{ij}$ do not depend on the coordinate $q^i$, then the $i$-th component $p_i=g_{ij}\dot{q}^j$ of the conjugate momentum is conserved on the trajectory.

\section{Geometric Formulation of Newtonian Dynamics}\label{gnd}

We begin our derivation by applying the geometric approach to classical Newtonian dynamics. Based on the discussion in the previous section, we replace the coordinate time $t$ (the classical evolution parameter)  by the proper time $\tau$. In this modification, Newton's second law for a force with potential $U$  becomes
\begin{equation}\label{NewtonSecondLawIND}
m\frac{d^2\mathbf{x}}{d\tau^2}=-\nabla U.
\end{equation}
Taking the Euclidean dot product of both sides of (\ref{NewtonSecondLawIND}) with $\dot{\mathbf{x}}$ gives
\begin{displaymath}
m\ddot{\mathbf{x}}\cdot\dot{\mathbf{x}}=-\nabla U\cdot\dot{\mathbf{x}},
\end{displaymath}
which, upon integration, yields
\begin{equation}\label{EnergyConserIND}
   \frac{m\dot{\mathbf{x}}^2}{2}+U(\mathbf{x})=E,
\end{equation}
where the integral of motion $E$ is the total energy on the object's worldline. The only difference between equation (\ref{EnergyConserIND}) and the classical energy conservation equation is the kinetic energy term, in which $d\mathbf{x}/dt$ has been replaced by $\dot{\mathbf{x}}$. In the classical limit, we have $dt= d\tau$, and then equation (\ref{EnergyConserIND}) reduces to the classical energy conservation equation.

Assume now that the potential $U(\mathbf{x})\leq 0$ and vanishes at infinity. Introduce the \textit{dimensionless potential}
\begin{equation}\label{u_def}
u(\mathbf{x})=\frac{-2U(\mathbf{x})}{mc^2},
\end{equation}
where $c$ denotes the speed of light. With this definition, equation (\ref{EnergyConserIND}) yields the dimensionless energy conservation equation
\begin{equation}\label{clenConsIND}
\frac{\dot{\mathbf{x}}^2}{c^2}-u=\mathcal{E},
\end{equation}
where $\mathcal{E}$ denotes the dimensionless total energy on the worldline. The total energy is a sum of kinetic energy, depending on the magnitude of the velocity, and potential energy, depending on position.

We turn now to the construction of the metric of the object's spacetime, for motion satisfying (\ref{NewtonSecondLawIND}), where $U$ is the potential of a static force. In our inertial lab frame, the metric is of the form
\begin{equation}\label{genmetricIND}
ds^2=f(\mathbf{x})c^2dt^2-g(\mathbf{x})d\mathbf{x}^2,
\end{equation}
where $f(\mathbf{x})$ and $g(\mathbf{x})$ depend solely on $\mathbf{x}$. Note that there are no anisotropic terms in the metric because in Newtonian dynamics, space is isotropic. Moreover, assuming Einstein synchrony, a straightforward argument shows that there are no time-space cross terms (see \cite{Rindler}, page 187).

For $u(\mathbf{x})\ll 1$, the worldlines are approximately straight, implying that this metric is asymptotically Minkowski. Hence,
 $f(\mathbf{x})\rightarrow 1$ and $g(\mathbf{x})\rightarrow 1$ as $u(\mathbf{x})\rightarrow 0$.

Since the metric is static, Proposition 1 implies that the zero component of the conjugate momentum is conserved. Thus,
\begin{equation}\label{Kcond2IND}
f(\mathbf{x})\dot{t}=k \quad,\quad \dot{t}=\frac{k}{f(\mathbf{x})},
\end{equation}
for some constant $k$ related to the total energy on the worldline.
The square of the norm with respect to (\ref{genmetricIND}) of the four-velocity $\left(\dot{t},\dot{\mathbf{x}}\right)$  is
\begin{displaymath}
f(\mathbf{x})c^2\dot{t}^2-g(\mathbf{x})\dot{\mathbf{x}}^2=c^2,
\end{displaymath}
which, by the use of (\ref{Kcond2IND}), leads to
\begin{equation}\label{4VCons}
\frac{k^2}{f(\mathbf{x})}-g(\mathbf{x})\frac{\dot{\mathbf{x}}^2}{c^2}=1.
\end{equation}
This can be considered a conservation equation on the worldline.

We can now determine the metric coefficients $f(\mathbf{x})$ and $g(\mathbf{x})$ by comparing this conservation to the energy conservation. Adding $1$ to both sides of equation (\ref{clenConsIND}) and dividing by $-(1-u)$, we obtain
\begin{equation}\label{normeqn3IND}
\frac{1}{1-u}\left(\mathcal{E}+1-\frac{\dot{\mathbf{x}}^2}{ c^2}\right)=1.
\end{equation}
Comparing (\ref{4VCons}) and (\ref{normeqn3IND}), and using  $f(\mathbf{x})\rightarrow 1$ as $u\rightarrow 0$, one obtains
\begin{equation}\label{gis}
g(\mathbf{x})=\frac{1}{1-u}\quad,\quad f(\mathbf{x})=1-u\quad,\quad k=\sqrt{\mathcal{E}+1}.
\end{equation}
From (\ref{genmetricIND}) and (\ref{gis}), we obtain the \textit{Newtonian metric}
\begin{equation}\label{metric}
ds^2=(1-u(\mathbf{x}))c^2dt^2-\frac{1}{1-u(\mathbf{x})}d\mathbf{x}^2.
\end{equation}
Reversing our argument shows that a trajectory which is minimal with respect to this metric satisfies Newton's second law (\ref{NewtonSecondLawIND}).

In order to complete the spacetime description of the worldline, from (\ref{Kcond2IND}) and (\ref{gis}) we obtain
\begin{equation}\label{ttau}
\dot{t}=\frac{k}{1-u(\mathbf{x})}=\frac{\sqrt{\mathcal{E}+1}}{1-u(\mathbf{x})}.
\end{equation}

\section{The deficiency of the Newtonian metric and the corrected metric}\label{refnewmet}

The huge success of Newtonian dynamics implies that the Newtonian metric (\ref{metric}) is close to the one that governs the laws of Nature. Nevertheless, the observed astrophysical deviations from the predictions of this dynamics indicate that this metric has a deficiency and needs to be corrected.

It is natural to assume that time intervals are influenced by the potential and should be altered by a factor defined by this potential when translated to the inertial frame. This influence is handled by the coefficient
$1-u$ of $c^2dt^2$ in (\ref{metric}) and accurately predicts the known gravitational time dilation in a spherically symmetric gravitational field. It is also natural to assume that the space increments in the direction of the gradient $\nabla U$
are influenced by the potential and should be altered by a factor defined by this potential when translated to the inertial frame. This influence is also present in (\ref{metric}).

However, the metric (\ref{metric}) is deficient in that it is {isotropic} - it alters the spatial increments equally in \emph{all} spatial directions. The potential, on the other hand, influences only the direction of the gradient $\nabla U$ and has no influence on the spatial increments in the directions transverse to the gradient. To remove this problem, we alter the metric so that the potential affects \emph{only} the direction of the force, in the same way as is in (\ref{metric}), and leaves the transverse directions unaffected.

More precisely, introduce at each $\mathbf{x}$ where $\nabla U(\mathbf{x})\neq \mathbf{0}$  a normalized vector
  \begin{equation}\label{n-def}
 \mathbf{n}(\mathbf{x}) = \frac{\nabla U(\mathbf{x})}{|\nabla U(\mathbf{x})|}
\end{equation}
 in the direction of the gradient of $U(\mathbf{x})$, or the negative of the direction of the force. Let $d\mathbf{x}_n=(d\mathbf{x}\cdot\mathbf{n})\mathbf{n}$ and  $d\mathbf{x}_{tr}=d\mathbf{x}-(d\mathbf{x}\cdot\mathbf{n})\mathbf{n}$,  respectively, denote the projections of the spatial increment $d\mathbf{x}$ in the parallel and transverse directions to $\mathbf{n}(\mathbf{x})$. With this notation, the \textit{corrected Newtonian metric} is
 \begin{equation}\label{normRND}
 ds^2=(1-u(\mathbf{x}))c^2dt^2-\frac{1}{1-u(\mathbf{x})} d\mathbf{x}_n^2-d\mathbf{x}_{tr}^2.
\end{equation}
We will call the dynamics resulting from this metric \textit{Relativistic Newtonian Dynamics} ($RND$).

At the points $\mathbf{x}_0=x_0^j$ where $\nabla U(\mathbf{x}_0)= \mathbf{0}$, the normalized vector $\mathbf{n}$ is not defined.   We claim that on any smooth trajectory $q(\sigma)$ in spacetime with  $q(\sigma_0)=(x_0^0,\mathbf{x}_0)=x_0$, the metric (\ref{normRND})  can be extended continuously to the point $x_0$. The Taylor expansion of  the potential $U(\mathbf{x})$ at  $\mathbf{x}_0$ to second order is $U(\mathbf{x})\approx U(\mathbf{x}_0)+\frac{1}{2} U_{,jk}(\mathbf{x}_0)(x^j-x_0^j)(x^k-x_0^k)$. The limit of  $\mathbf{n}(\sigma)$ as $\Delta\sigma=\sigma-\sigma_0\rightarrow 0$ on the trajectory $q(\sigma)$ can be calculated by the limit along its tangent line $x_0^\alpha +a^\alpha \Delta\sigma$, where $a$ is the tangent vector to the trajectory at $x_0$, and use of the second-order approximation of $U(\mathbf{x})$ to calculate  $\nabla U(\mathbf{x})$. This yields
$\nabla U(\mathbf{x}(\Delta\sigma))_k=\frac{1}{2} U_{,jk}(\mathbf{x}_0)(a^j\Delta\sigma)=\mathbf{b}\Delta\sigma$. Hence,
\[\lim_{\Delta\sigma\to 0+} \mathbf{n}(\Delta\sigma)=-\lim_{\Delta\sigma\to 0-} \mathbf{n}(\Delta\sigma).\]
Since the metric is not affected by the sign of  $\mathbf{n}$, this proves our claim. Moreover, since, in general, the measure of such points $\mathbf{x}_0$ is zero, the length of the trajectory is not affected by the metric at these points.

In the case of the gravitational field of a non-rotating, spherically symmetric body of mass $M$, in spherical coordinates with origin at its center, the potential is $U(r)=-GmM/r$, and the dimensionless  potential (\ref{u_def}) is $u(r)=2GM/c^2r=r_s /r$, where $r_s=2GM/c^2$ is the Schwarzschild radius. In this case, the metric (\ref{normRND}) is
\begin{equation}\label{Schwatz}
   ds^2=\left(1-\frac{r_s}{r} \right)c^2dt^2-\frac{1}{1-r_s/r} dr^2-r^2d\theta^2-r^2\sin ^2\theta d\varphi^2,
\end{equation}
which is the well-known Schwarzschild metric (\cite{MTW}).

\section{The Equations of Relativistic Newtonian Dynamics}\label{rnd}

We now obtain the dimensionless and dimensional \textit{energy conservation equations} and \textit{equations of motion} of $RND$. The derivation is similar to the reversal of the derivation in Section \ref{gnd} for the metric (\ref{metric}). Since the metric (\ref{normRND}) is static, Proposition 1 implies here, as in (\ref{Kcond2IND}), that
\begin{equation}\label{Kcond3IND}
\dot{t}=\frac{k}{1-u(\mathbf{x})}.
\end{equation}
The square of the norm with respect to (\ref{normRND}) of the four-velocity $\dot{x}$ is
\begin{equation}\label{RNDnormEqn}
c^2\frac{k^2}{1-u}-\frac{1}{1-u}\dot{\mathbf{x}}_n^2-\dot{\mathbf{x}}_{tr}^2=c^2.
\end{equation}
Multiplying by $\frac{1-u}{c^2}$, using $\dot{\mathbf{x}}^2=\dot{\mathbf{x}}_n^2+\dot{\mathbf{x}}_{tr}^2$ and rearranging terms, we obtain the  \textit{dimensionless energy conservation equation}
\begin{equation}\label{clenConsRND}
\frac{\dot{\mathbf{x}}^2}{c^2}-u-u\frac{\dot{\mathbf{x}}_{tr}^2}{c^2}=k^2-1.
\end{equation}
The corresponding  \textit{dimensional energy conservation equation} is
\begin{equation}\label{RNDenergyCos}
   \frac{m\dot{\mathbf{x}}^2}{2}+U(\mathbf{x}) +U(\mathbf{x})\frac{\dot{\mathbf{x}}_{tr}^2}{c^2}=E,
\end{equation}
where the integral of motion $E$ is the total energy on the worldline. As in the energy conservation equation (\ref{EnergyConserIND}) of modified Newtonian dynamics, equation (\ref{RNDenergyCos}) has a kinetic energy term and a potential energy term. But in (\ref{RNDenergyCos}), there is also a \emph{mixed term} which depends on both the velocity of the object and the potential. This means that in order to reproduce relativistic effects, one can no longer distinguish between potential and kinetic energy, as in Newtonian dynamics. This also explains the need to include the velocity in the modified Newtonian potentials proposed in \cite{TR,Ghosh14,SG,G15,Ghosh16}. The mixed term in (\ref{RNDenergyCos}) is approximately $\beta^2U(\mathbf{x})$ and is therefore only seen for high velocities or in high-precision experiments.

Let $\phi=U/m$ denote the potential per unit mass. Differentiating equation (\ref{RNDenergyCos}) with respect to $\tau$, one obtains the \textit{equation of motion} of $RND$
\begin{equation}\label{RND1finals}
 \ddot{\mathbf{x}}=-\nabla \phi-\nabla \phi\frac{\dot{\mathbf{x}}^2_{tr}}{c^2}+ 2\frac{\phi(\mathbf{x})}{c^2}(\dot{\mathbf{x}}\cdot\dot{\mathbf{n}})\mathbf{n},
\end{equation}
which has now two additional terms not appearing in the corresponding classical equation (\ref{NewtonSecondLawIND}). In the classical regime, both of these terms are small and have therefore gone unrecognized.

For potentials such as the gravitational potential, for which $\phi$ is independent of $m$,  equations (\ref{clenConsRND}) and (\ref{RND1finals}) can be extended to massless particles as well by using the
symbol $\varepsilon$, which equals $1$ for \textit{objects} with non-zero mass and $0$ for massless \textit{particles}.  However, for massless particles, the proper time $\tau$ is not defined. Instead, we will use an affine parameter (see,
for example, \cite{MTW}). There is no need here to specify this parameter, because to test our theory, we obtain parameter-free equations.

For massless particles, the norm of the four-velocity $\dot{x}=\left(\dot{t},\dot{\mathbf{x}}\right)$ is $0$.
Replacing $c^2$ by $0$ on the right-hand side of equation (\ref{RNDnormEqn}), we obtain
 \begin{equation}\label{ConsRNDgen}
\frac{\dot{\mathbf{x}}_n^2}{c^2}+(1-u)\left(\frac{\dot{\mathbf{x}}_{tr}^2}{c^2}+\varepsilon\right)=k^2.
\end{equation}
Differentiating with respect to $\tau$ yields, in turn,
\begin{equation}\label{eomRND}
 \ddot{\mathbf{x}}=-\varepsilon\nabla \phi-\nabla \phi\frac{\dot{\mathbf{x}}^2_{tr}}{c^2}+ 2\frac{\phi(\mathbf{x})}{c^2}(\dot{\mathbf{x}}\cdot\dot{\mathbf{n}})\mathbf{n},
\end{equation}
{which is valid everywhere except on the Schwarzschild horizon.} Equations (\ref{ConsRNDgen}) and (\ref{eomRND}) are, respectively, the \emph{dimensionless energy conservation equation} and the \emph{dimensionless equation of motion for objects/particles} in $RND$. Equations (\ref{Kcond3IND}) and (\ref{eomRND}) provide a complete description of a worldline in $RND$.

Note that even though the classical force is zero for a massless particle, the second and third terms in the equation of motion (\ref{eomRND}) nevertheless remain and account for phenomena such as gravitational lensing and the Shapiro time delay.

{It is clear from these equations that $RND$ reduces in the low velocity, weak field limit to classical Newtonian dynamics. We show in the next section that for a gravitational potential, this
dynamics exactly reproduces the classical tests of $GR$.}

\section{Tests of $GR$ explained with $RND$}\label{tests}

For the gravitational field of a non-rotating, spherically symmetric body of mass $M$, the unit vector $\mathbf{n}$, defined by (\ref{n-def}), is the radial direction, so $\dot{\mathbf{x}}_n=\dot{r}$. If the initial position and velocity of the object/particle are in the plane $\theta=\pi/2$, then they will remain in this plane throughout the motion (\cite{Gron}). Thus, one may chose the coordinate system so that $\theta=\pi/2$.

Moreover, the metric coefficients (\ref{Schwatz}) are independent of $\varphi$. Hence, by Proposition 1, we have
\begin{equation}\label{AngMom}
r^2\dot{\varphi}=J,
\end{equation}
where, for objects with non-zero mass, the constant $J$ has the meaning of angular momentum per unit mass. This implies that $\dot{\mathbf{x}}^2_{tr}=r^2\dot{\varphi}^2=\frac{J^2}{r^2}$, and one can rewrite equation  (\ref{ConsRNDgen}) as
\begin{equation}
  \frac{\dot{r}^2}{c^2} +(1-u(r))\left(\frac{J^2}{c^2r^2}+\varepsilon \right)=k^2.
\end{equation}
This, together with the definition of $J$, leads to the \emph{path equation for a central force}
\begin{equation}\label{Trajcen}
  \left(\frac{J}{cr^2}\frac{dr}{d\varphi}\right)^2 +(1-u(r))\left(\frac{J^2}{c^2r^2}+\varepsilon \right)=k^2,
\end{equation}
which depends on the two integrals of motion $k$ and $J$ and coincides with the geodesic equation of the Schwarzschild metric (\cite{MTW}).

Furthermore, from (\ref{Kcond3IND}) and (\ref{AngMom}), the \emph{time dependence equation for a central force} on this path is
\begin{equation}\label{Timetr}
  \frac{dt}{d\varphi}=\frac{kr^2}{J(1-u(r))}.
\end{equation}

For a non-rotating, spherically symmetric object of mass $M$, the dimensionless gravitational potential defined by (\ref{u_def}) is $u(r)=\frac{r_s}{r}$, where $r_s= \frac{2GM}{c^2}$ is its Schwarzschild radius.

\textbf{1.} \quad To describe the trajectory of Mercury ($\varepsilon=1$) in the gravitational field of the Sun, we rewrite (\ref{Trajcen}) in terms of the dimensionless potential energy $u$ by substituting $r=r_s/u$. Defining the orbit parameter $\mu=\frac{1}{2}\left(\frac{cr_s}{J}\right)^2$, we obtain the \emph{$RND$ equation for the planetary orbit}
\begin{equation}\label{TrajPlanet}\left(\frac{du}{d\varphi}\right)^2= u^3-u^2+2\mu u +2\mu(k^2-1),\end{equation}
which is similar to the corresponding equation in $GR$ (see, for example, \cite{Bro}).

The corresponding classical Newtonian equation for this orbit is
\begin{equation}\left(\frac{du}{d\varphi}\right)^2= -u^2+2\mu u +2\mu(k^2-1).\end{equation}
For a bounded orbit,  the maximum and minimum values of $u$ are the roots $u_p, u_a$ of the quadratic on the right-hand side of this equation, corresponding to the perigee and apogee, respectively. This equation has the obvious classical solution $u_{cl}(\varphi)=\mu(1+e\cos(\varphi-\varphi_0))$, where $\varphi_0$ is the polar angle of the perigee and $e$ is the eccentricity of the orbit. Here, $u_p=\mu(1+e)$ and $u_a=\mu(1-e)$. Then $\mu=(u_p+u_a)/2$ is the average energy on the trajectory.  In polar coordinates, we have
\begin{equation}r_{cl}(\varphi)=\frac{r_s/\mu}{1+e\cos(\varphi-\varphi_0)},\end{equation}
which is a non-precessing ellipse. The reason there is no precession is that the radial and angular periodicities are both equal to $2\pi$. This is no longer the case in $RND$ dynamics, {due to the anisotropy of the metric (\ref{Schwatz}).}


The $RND$ solution (the solution of (\ref{TrajPlanet})) is of the form $u(\varphi)=\mu(1+e\cos\alpha(\varphi))$, where the angle $\alpha$ satisfies $r_{cl}(\alpha)=r(\varphi)$. As in the classical case, $u_p, u_a$ are again roots of the cubic on the right-hand side of equation (\ref{TrajPlanet}), but this cubic has an additional root $u_3=1-(u_p+u_a)=1-2\mu$.  Substituting this into (\ref{TrajPlanet}), one obtains $d\alpha/d\varphi =\sqrt{1-3\mu-\mu e\cos\alpha}$ and the explicit dependence
\begin{equation} \varphi(\alpha) =\varphi_0+\int_0^\alpha (1-3\mu-\mu e\cos\tilde{\alpha})^{-1/2}d\tilde{\alpha},\end{equation}
which eventually yields the known perihelion precession formula (\cite{MTW})
\begin{equation}\label{pecM}
  \varphi(2\pi)-\varphi(0)-2\pi\approx 3\pi \mu \frac{rad}{rev}.
\end{equation}
Substituting the value of $\mu$ for Mercury, we obtain its observed anomalous precession.

\textbf{2.} \quad Since, in $RND$, we use the unmodified Newtonian potential, the potential of a binary star is the same as the potential of a classical two-body problem. Therefore, we can reduce the problem to a one-body problem in the gravitational field of an object with mass $M$, the combined mass of the binary, located at the center of mass of the binary. Hence, the $RND$ treatment of the binary is the same as for Mercury and will once again produce precessing elliptic orbits for each component of the binary, with precession given by (\ref{pecM}).  As shown in \cite{FSBin}, this leads to a periastron advance
\begin{equation}\label{BinPrec}
  \dot{\omega} =3\frac{(GM)^{2/3}}{c^2(1-e^2)}\left(\frac{P_b}{2\pi}\right)^{-5/3},
\end{equation}
where $P_b$ is the orbital period of the binary and $\omega$ is the angular position of the periastron. This formula is identical to the post-Keplerian formula for the relativistic advance of the periastron found in \cite{DD}.

\textbf{3.} \quad  Gravitational lensing and the Shapiro time delay (or gravitational time delay) describe the deflection of a light ray and the slowing of a light pulse ($\varepsilon=0$) as it moves from a point $A$ to a point $B$  in the gravitational potential of a spherically symmetric massive object of mass $M$. Denote by $r_0$ the distance from the point $P$ on the trajectory closest to the massive object. Since $\frac{dr}{d\varphi}=0$ at the point $P$, it follows from (\ref{Trajcen}) that
\begin{equation}\label{Jk}\frac{J^2}{c^2k^2}=-\frac{r_0^2}{1-r_s /r_0}.\end{equation}

To obtain the formula for gravitational lensing, substitute (\ref{Jk}) into (\ref{Trajcen}), which yields
\begin{equation}\left(\frac{r_0}{r^2}\frac{dr}{d\varphi}\right)^2+\left(1-\frac{r_s}{r}\right)\frac{r_0^2}{r^2}=1-\frac{r_s}{r_0}.\end{equation}
For any angle $\varphi$ on the trajectory, one may associate an angle $\alpha(\varphi)$ for which $r(\varphi)=\bar{r}(\alpha)$, where $\bar{r}(\alpha)=\frac{r_0}{\sin \alpha}$ is the
straight-line approximation of the trajectory at the point $P$, chosen to be the $x$ direction.
This suggests the substitution $r=\frac{r_0}{\sin\alpha}$,  which implies $\frac{dr}{d\varphi}=-\cos\alpha \frac{r^2}{r_0}\frac{d\alpha}{d\varphi}$ and
\begin{equation}\label{dpda}\frac{d\varphi}{d\alpha}=\left( 1-\frac{r_s}{r_0}\left(\sin \alpha+\frac{1}{1+\sin\alpha}\right)\right)^{-1/2}.\end{equation}
Hence, the deflection angle of a light ray moving from $A$ to $B$ is
\begin{equation}\delta\phi=\int_{\alpha_A}^{\alpha_B}\left(1-\frac{r_s}{r_0}\left(\sin \alpha+\frac{1}{1+\sin\alpha}\right)\right)^{-1/2}d\alpha-\pi,\end{equation}
where $\alpha_A, \alpha_B$ are the $\alpha$ values of the points $A$ and $B$, respectively.
Assuming that these points are very remote from the massive body ($\alpha_A\approx \pi, \alpha_B\approx 0$) and that $r_s/r_0\ll 1$, the weak deflection angle becomes
\begin{equation}\label{Lens}
 \delta\phi\approx\frac{2r_s}{r_0}=\frac{4GM}{c^2 r_0},
\end{equation}
which is identical to the angle given by Einstein's formula for weak gravitational lensing using $GR$ (\cite{MTW,CW}).

\textbf{4.} \quad To obtain the formula for the Shapiro time delay, one substitutes the value of $J/ck$ from (\ref{Jk}) into the $RND$ time dependence equation for a central force (\ref{Timetr}). Hence, the time of passage of light from the point $P$ to $B$ is given by
\begin{equation}\label{tb}c(T_B-T_P)=\int_{\varphi_B}^{\pi/2}\frac{r^2\sqrt{1-r_s /r_0}}{(1-r_s/r)r_0}d\varphi \,.\end{equation}
For the common case $r_s/r_0\ll 1$, we work in first order in $r_s/r_0$. Then, using (\ref{dpda}) and the same substitution $r=\frac{r_0}{\sin\alpha}$ as above,
the time propagation between $P$ and $B$ is
\begin{equation}\label{TimeDe}
c(T_B-T_P)\approx x_B+r_s\ln \frac{r_B+x_B}{r_0},
\end{equation}
where  $x_B$  denotes the $x$ coordinate of $B$. Using this approximation, the Shapiro time delay for a signal traveling from
$A$ to $B$ and back is
\begin{equation}\label{TimeDelf}
  r_s\ln \frac{4x_B|x_A|}{r_0^2},
\end{equation}
which is the known formula for the Shapiro time delay (\cite{MTW,CW}), confirmed by several experiments.

\section{Discussion}\label{disc}

Riemann's approach to unify the laws of electricity, magnetism, light and gravitation was geometric. He believed that the forces at play in a system determine the geometry of the system. Put simply, for Riemann, \emph{force equals geometry}. His quest failed, unfortunately, because he considered how forces affect \emph{space}, not \emph{spacetime}. Nevertheless, his geometric approach led to the development of \emph{pseudo-Riemannian geometry} which fifty years later provided the cornerstone for Einstein's $GR$. However, $GR$ singles out the gravitational force from other forces which are not treated geometrically.

In this paper we introduced the relativity of spacetime in order  to apply Riemann's principle of ``force equals geometry" to the dynamics under  any static, conservative force. We accomplished this by describing the geometry of the spacetime of a moving object via a metric derived from the potential of the force field acting on the object. Since an inanimate object has no internal mechanism with which to change its velocity,   it has constant velocity in \emph{its} own world. This led us to formulate our new \emph{Generalized Principle of Inertia}, which states that: \textbf{An inanimate object moves inertially, that is, with constant velocity, in \emph{its own} spacetime whose geometry is determined by the forces affecting  it}.

This is a generalization, or more accurately, a relativization of what Einstein accomplished. In $GR$, an object freely falling in a gravitational field is in free motion. Along a geodesic, the acceleration is zero. The Generalized Principle of Inertia states that \emph{every} object is in free motion in \emph{its} own world, determined by the forces which affect it. Thus, we assumed the existence of a metric with respect to which the length of the object's trajectory is extremal, enabling us to use a variational principle and conserved quantities to calculate trajectories.

Specifically, we began by treating classical Newtonian dynamics within this framework, using a properly defined \emph{Newtonian metric}.
Nevertheless, our Newtonian metric is still deficient, since it fails to reproduce the tests of $GR$.  The deficiency lies in the fact that this metric is isotropic, while the potential influences only the direction of the force and has no influence on directions transverse to the force. We removed this deficiency and obtained a corrected Newtonian metric (\ref{normRND}). The dynamics built on the corrected metric is called \emph{Relativistic Newtonian Dynamics (RND)}. We derived the \textit{dimensionless energy conservation equation} (\ref{ConsRNDgen}) and the \textit{dimensionless equation of motion} (\ref{eomRND}) of $RND$, for both massless particles and objects with non-zero mass.

It is clear from these equations that this dynamics reduces in the low velocity, weak field limit to classical Newtonian dynamics. Moreover, as a partial validation of our approach, we have shown in section \ref{tests} that for a gravitational potential, $RND$ exactly reproduces the tests of $GR$. The derivation of our metric is much simpler than in $GR$ and uses  potentials defined by the sources via Poisson's equation for the static case. We expect $RND$ to be useful for studying relativistic gravitational astrophysical (or other) phenomena.

In classical physics, the total energy has two mutually exclusive contributions: the kinetic energy depending only on the magnitude of the velocity of the object, and the potential energy depending on the field and the object's position. This leads to the isotropy of the spatial part of the Newtonian metric (\ref{metric}). In $RND$, on the other hand, the metric is not spatially isotropic. This is reflected in the additional term
\begin{displaymath}\label{mixedterm}
U(\mathbf{x})\frac{\dot{\mathbf{x}}_{tr}^2}{c^2}
\end{displaymath}
of the energy conservation equation (\ref{RNDenergyCos}), which contains contributions from both kinetic and potential energy.  This implies that in order to reproduce relativistic effects, one can no longer separate these contributions. Indeed, some authors  \cite{TR,Ghosh14,SG,G15,Ghosh16}  have defined velocity-dependent ``potentials'' in order to reproduce relativistic effects. However, these potentials are not potentials in the true sense. They are ``Newtonian analogous potentials" (see \cite{Ghosh16}).

The $RND$ model in its present nascent form is restricted to a static conservative force field. In the case of the gravitational field of a non-rotating, spherically symmetric body, in spherical coordinates with origin at its center, the corrected Newtonian metric (\ref{Schwatz}) underlying the $RND$ model reduces to the well-known Schwarzschild metric. Hence, of the ten post-Newtonian parametrization (PPN) parameters characterizing the weak-field behavior of a metric theory, the $RND$ model is characterized by the only two non-zero $PPN$ parameters, $\beta=\gamma=1$.  These parameters, measuring the nonlinearity in the superposition law for gravity and the space-curvature produced by unit rest mass, respectively, are sufficient to describe the classical tests of $GR$.

At this stage, the $RND$ model does not describe modern tests of $GR$ beyond the classical tests. In particular, since the potential of a collapsing binary is not static, the model in its current form does not provide a mechanism for the recently discovered gravitational waves by the LIGO team. We hope to extend the model to handle the modern tests as well.

By applying Lorentz covariance to the static case, we also hope to extend the model  for fields generated by moving sources. This could be achieved by extending Mashhoon’s linear perturbation approach of gravitoelectromagnetism \cite{Mash} to higher order. This is necessitated by the fact that (as seen from equation (\ref{eomRND} )), the relativistic corrections to the force in the $RND$ model are of second order in $v/c$, whereas in Mashhoon’s approach they are of first order in $v/c$.  In the resulting extended model we will be able to calculate the remaining eight $PPN$ parameters in order to evaluate the model.

 Kepler's laws of planetary motion in celestial mechanics provided the basis for Newtonian physics, applicable until today to all forces of Nature in the non-relativistic regime. In a similar way, we expect $RND$ to provide the basis for relativistic physics for other forces of Nature.
\begin{acknowledgements}
 We  wish  to thank the referee for his constructive comments.
\end{acknowledgements}

\end{document}